\documentclass[aps, reprint,10pt,twocolumn,superscriptaddress]{revtex4-1}
\usepackage{amsmath}
\usepackage{latexsym}
\usepackage{amssymb}
\usepackage{graphics,epstopdf}
\usepackage{hyperref}
\hypersetup{
    colorlinks = true,
    linkcolor =blue,
	citecolor=blue, 
	urlcolor=blue 
}

\newcommand{\ed}{\end{document}}
\newcommand{\beq}{\begin{equation}}
\newcommand{\eeq}{\end{equation}}

\usepackage{mathtools}
\usepackage{natbib}

\usepackage[normalem]{ulem}

\begin{document}
\title{Universal subdiffusive behavior at band edges from transfer matrix exceptional points}

\author{Madhumita Saha}
\email{madhumita.saha@acads.iiserpune.ac.in} 
\affiliation{Department of Physics, Indian Institute of Science Education and Research Pune, Dr. Homi Bhabha Road, Ward No. 8, NCL Colony, Pashan, Pune, Maharashtra 411008, India}
\affiliation{International Centre for Theoretical Sciences, Tata Institute of Fundamental Research, Bangalore 560089, India}

\author{Bijay Kumar Agarwalla}
\email{bijay@iiserpune.ac.in}
\affiliation{Department of Physics, Indian Institute of Science Education and Research Pune, Dr. Homi Bhabha Road, Ward No. 8, NCL Colony, Pashan, Pune, Maharashtra 411008, India}

\author{Manas Kulkarni}
	\email{manas.kulkarni@icts.res.in}
	\affiliation{International Centre for Theoretical Sciences, Tata Institute of Fundamental Research, Bangalore 560089, India}
\author{Archak Purkayastha}
\email{archak.p@phys.au.dk}
\affiliation{School of Physics, Trinity College Dublin, Dublin 2, Ireland}
\affiliation{Center for Complex Quantum Systems, Department of Physics and Astronomy, Aarhus University, Ny Munkegade 120, DK-8000 Aarhus C, Denmark.}
\affiliation{Department of Physics, Indian Institute of Technology, Hyderabad 502284, India}
\date{\today} 

\begin{abstract}
We discover a deep connection between parity-time (PT) symmetric optical systems and quantum transport in one-dimensional fermionic chains in a two-terminal open system setting. The spectrum of one dimensional  tight-binding chain with periodic on-site potential can be obtained by casting the problem in terms of $2 \times 2$ transfer matrices. We find that these non-Hermitian matrices have a symmetry exactly analogous to the PT-symmetry of balanced-gain-loss optical systems, and hence show analogous transitions across exceptional points. We show that the exceptional points of the transfer matrix of a unit cell correspond to the band edges of the spectrum. When connected to two zero temperature baths at two ends, this consequently leads to subdiffusive scaling of conductance with system size, with an exponent $2$, if the chemical potential of the baths are equal to the band edges. We further demonstrate the existence of a dissipative quantum phase transition as the chemical potential is tuned across any band edge. Remarkably, this feature is analogous to transition across a mobility edge in quasiperiodic systems. This behavior is universal, irrespective of the details of the periodic potential and the number of bands of the underlying lattice. It, however, has no analog in  absence of the baths.
\end{abstract}

\maketitle

{\it Introduction and overview of results ---}
It is fascinating how fundamental mathematical concepts aid  in our understanding of physical phenomena across all scales. This often allows us to find deep connections between seemingly completely disparate physical situations. Here, we show how a property termed  pseudo-Hermiticity of non-Hermitian matrices reveal a unique connection between balanced-gain-loss optical systems and quantum transport in fermionic chains.   

The dynamics of two coupled optical cavities, one with gain the other with loss, such that the gain and loss are perfectly balanced, is most commonly envisioned as being governed by the so-called $2 \times 2$ parity-time symmetric (PT) non-Hermitian `Hamiltonian' $\mathbf{H}_{\rm PT}=\omega_0\mathbb{I}_2 + i\gamma \sigma_z+g\sigma_x$ \cite{El_Ganainy2019,El_Ganainy2018,Feng2017}. Here $\sigma_{x,y,z}$ are the Pauli matrices, and $\mathbb{I}_2$ is the $2 \times 2$ identity matrix. This $2 \times 2$ non-Hermitian matrix has the pseudo-Hermiticity property associated with the antilinear operator $\sigma_x \mathcal{K}$, i.e, $\left(\sigma_x \mathcal{K}\right)\mathbf{H}_{\rm PT} \left(\sigma_x \mathcal{K}\right)^{-1}=\mathbf{H}_{\rm PT}$, where $\sigma_x$ describes the parity operator and $\mathcal{K}$ describes the time-reversal (complex conjugation) operator. Whenever a matrix has such an pseudo-Hermiticity, its eigenvalues are either purely real or occur in complex conjugate pairs \cite{Wigner_1960,Bender_1998,Mostafazadeh_2002}. When the eigenvalues are real, the eigenvectors are also simultaneous eigenvectors of the symmetry operator $\sigma_x \mathcal{K}$ with eigenvalue $1$. This is termed PT-symmetric regime. When the eigenvalues are complex, the eigenvectors of $\mathbf{H}_{\rm PT}$ are no longer simultaneous eigenvectors of the symmetry operator. 
This is termed  PT-broken regime.  Transition between these two regimes occurs at $\gamma=g$ which is the exceptional point (EP), where there is a single eigenvalue $\omega_0$ and the matrix is not diagonalizable. The dynamics drastically changes on transition across the EP, leading to interesting applications and exotic physics in both classical and quantum regimes \cite{El_Ganainy2019,El_Ganainy2018,Feng2017,Archak_2020_PT,Lau_2018,
Kumar_2022}. This is the most paradigmatic example of symmetries of non-Hermitian matrices governing physical systems \cite{Franca_2021,Wang_2019,Arkhipov_2021,Roccati_2021,Arkhipov_2020,
Gomez_Leon_2021,
Mcdonald_2021,Avila_2020,Nori_2018,Huber_2020,Kazuki_2019, Khandelwal_2021, Archak_2022, Abbasi_2022, Chen_2022, Kumar_2022, Roccati_2022,Ashida_2020,Bergholtz_2021, Ruzicka_2021,Kawabata_2019,Foa_Torres_2019,Gong_2018}.

 In this work, we explore the effects of a similar transition occurring in a different kind of non-Hermitian matrix that appears in scattering theory: the transfer matrix. Unlike mostly studied non-Hermitian matrices, transfer matrices do not directly govern the dynamics of the system. They instead play a fundamental role in determining the spectrum of the Hamiltonian of the system. The band-structure of nearest neighbour fermionic chains with periodic on-site potentials can be obtained by casting the problem in terms of $2 \times 2$ transfer matrices \cite{Last_1993,Molinari_1997,Molinari_1998}. We note that each such transfer matrix can be transformed to the form of $\mathbf{H}_{\rm PT}$ via a unitary transformation $\mathbf{U}$. Consequently, the transfer matrices have an pseudo-Hermiticity, associated with the antilinear operator  $S=\mathbf{U} \sigma_x \mathcal{K} \mathbf{U}^\dagger$. So, similar to $\mathbf{H}_{\rm PT}$, they show transitions across EPs from $S$-symmetric to $S$-symmetry-broken regimes.
 
\begin{figure}
    \centering
    \includegraphics[width=\columnwidth]{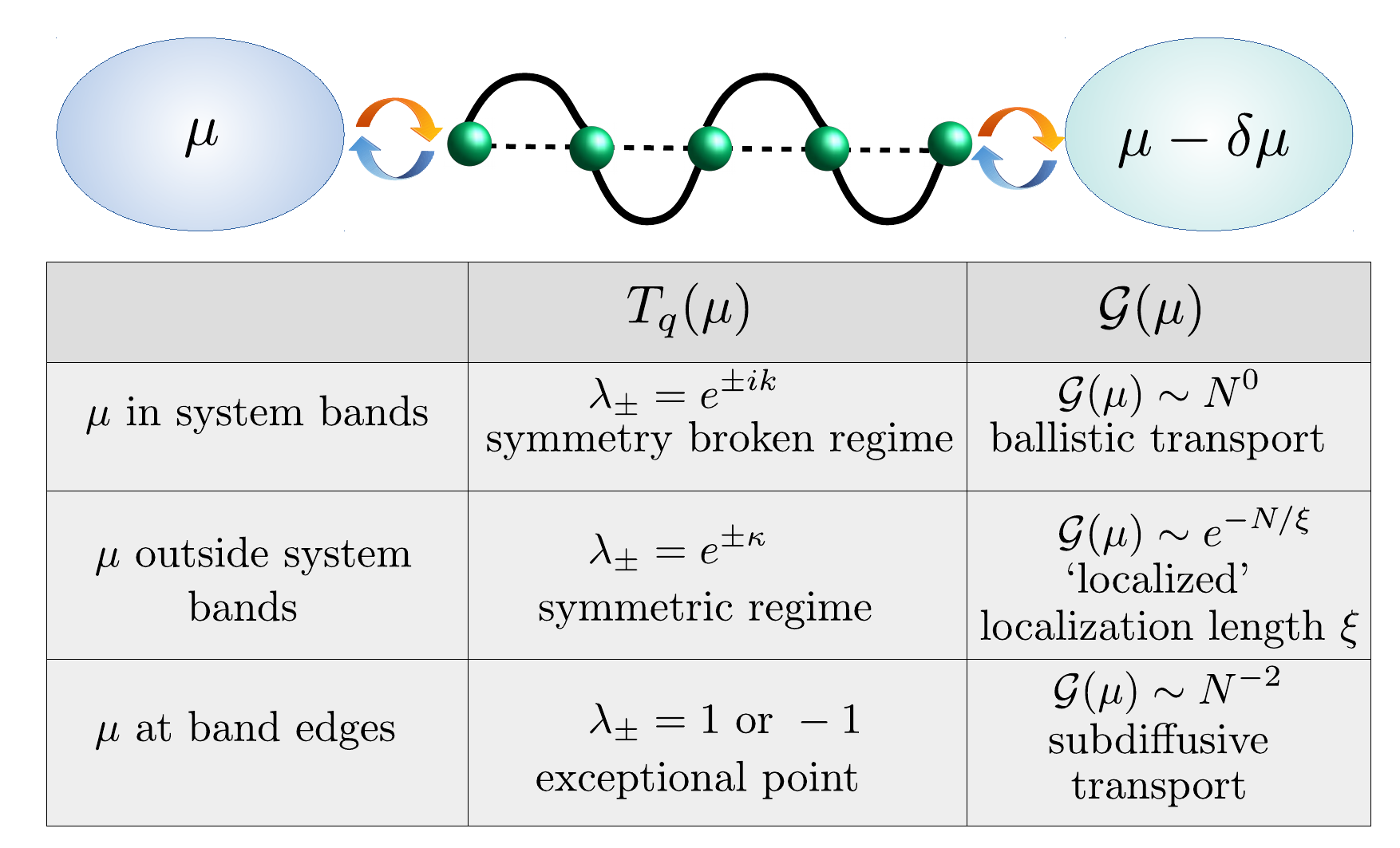}
    \caption{The top panel shows a schematic of a fermionic nearest neighbour hopping chain, with a periodic on-site potential coupled at two ends to two zero temperature baths at slightly different chemical potentials, $\mu$ and $\mu-\delta \mu$. The table summarizes our main result. Here $\mathbf{T}_{q}(\mu)$ is the transfer matrix of a unit cell, $\lambda_{\pm}$ are its eigenvalues, $\mathcal{G}(\mu)$ is the conductance in the two-terminal setting.} 
    \label{fig:1}
\end{figure}

We find that the EPs of the transfer matrix of a unit cell of the system correspond to the band edges. When the system is connected to two zero temperature baths at two ends (see Fig.~\ref{fig:1}), this, in turn, leads to a subdiffusive scaling of conductance with system size if the chemical potential $\mu$ is equal to any band edge. The subdiffusive scaling exponent is universal, irrespective of any further details of the periodic on-site potential. If $\mu$ is outside any system band, the eigenvalues of the transfer matrix are real ($S$-symmetric regime), which leads to lack of transport beyond a well-defined length scale. If $\mu$ is inside any system band, the eigenvalues of the transfer matrix are complex ($S$-symmetry-broken regime), which leads to ballistic transport. Thus, a transition across EP occurs in the transfer matrix when the chemical potential is tuned across a band edge. Correspondingly, there occurs a non-analytic change in the zero temperature steady state transport properties of the open system, thereby causing a dissipative  quantum phase transition. Our results can be summarized in Fig.~\ref{fig:1}. This transition occurring in the behaviour of conductance as a function of $\mu$ at every band edge is reminiscent of localization-delocalization transitions across a mobility edge occurring in certain one dimensional quasiperiodic systems (for example, \cite{DasSarma_1988,Luschen_2018,Ganeshan_2015,Li_2017, Wang_2020, Wang_2021}). We discuss the similarities and the differences between them.


{\it Tight-binding chain and transfer matrices ---}
We consider a fermionic nearest neighbour tight-binding chain of $N$ sites with a periodic potential,
\begin{equation}
\hat{H}_S = \sum_{\ell=1}^N \varepsilon_\ell \hat{c}_\ell^\dagger \hat{c}_{\ell}+ \sum_{\ell=1}^{N-1} (\hat{c}_\ell^\dagger \hat{c}_{\ell+1} + \hat{c}_{\ell+1}^\dagger \hat{c}_{\ell}),
\end{equation}
where $\hat{c}_\ell$ is the fermionic annihilation operator at site $\ell$ of the chain, and $\varepsilon_\ell$ is a periodic on-site potential satisfying $\varepsilon_{\ell+q}=\varepsilon_\ell$. Here $q$ is the length of the unit cell and the hopping parameter is set to $1$, which is therefore the unit of energy. We consider $N$ to be an integer multiple of $q$. The periodic on-site potential with unit cell of length $q$ causes the spectrum of the system to be separated into $q$ bands. In the thermodynamic limit, using Bloch's theorem, the energy dispersion of the bands can be obtained via solving the following equation for $\varepsilon$ \cite{Last_1993,Molinari_1997,Molinari_1998},
\begin{equation}
\label{band_condition}
{\rm Tr}\left(\mathbf{T}_q(\varepsilon)\right)=2 \cos k,
\end{equation}
where, $k$ is the wave-vector,  $-\pi\leq k \leq \pi$, and $\mathbf{T}_{q}(\varepsilon)$ is given by \cite{supp}
\begin{equation}
\label{def_T}
\mathbf{T}_{q}(\varepsilon) = \prod_{\ell=1}^q \mathbf{T}^{(\ell)}(\varepsilon),~\mathbf{T}^{(\ell)}(\varepsilon)=\frac{\varepsilon-\varepsilon_{\ell}}{2}\left(\mathbb{I}_2+\sigma_z\right)-i\sigma_y.
\end{equation}
Here, $\mathbf{T}^{(\ell)}(\varepsilon)$ is the transfer matrix for site $\ell$, whereas $\mathbf{T}_{q}(\varepsilon)$ is the transfer matrix for a single unit cell of the lattice. 

{\it Pseudo-Hermiticity of transfer matrices ---}
We carry out the following unitary transformation on the transfer matrix for site $\ell$,  $\mathbf{U}^\dagger \mathbf{T}^{(\ell)}(\varepsilon) \mathbf{U} = \frac{\varepsilon-\varepsilon_{\ell}}{2}\left(\mathbb{I}_2+\sigma_x\right)-i\sigma_z$, where $\mathbf{U}$ is the $2 \times 2$ unitary matrix that diagonalizes $\sigma_y$. The elements of $\mathbf{U}$ are $\mathbf{U}_{11}=1/\sqrt{2}$, $\mathbf{U}_{12}=1/\sqrt{2}$, $\mathbf{U}_{21}=i/\sqrt{2}$, $\mathbf{U}_{22}=-i/\sqrt{2}$. After the unitary transformation, $\mathbf{T}^{(\ell)}(\varepsilon)$ is of the exact same form as $\mathbf{H}_{\rm PT}$, and therefore commutes with $\sigma_x \mathcal{K}$. This, in turn means, $S\mathbf{T}^{(\ell)}(\varepsilon)S^{-1}=\mathbf{T}^{(\ell)}(\varepsilon)$, with $S=\mathbf{U}\sigma_x \mathcal{K} \mathbf{U}^\dagger$. Thus, transfer matrix for each site has the same pseudo-Hermiticity. Consequently, the transfer matrix of the unit cell, $\mathbf{T}_{q}(\varepsilon)$, which is obtained by multiplying transfer matrices for each site, also has the pseudo-Hermiticity associated with $S$.

The existence of the pseudo-Hermiticity guarantees that every transfer matrix has (a) a $S$-symmetric regime where eigenvalues are real, and eigenvectors are simultaneous eigenvectors of $S$, (b) a $S$-symmetry broken regime where the eigenvalues are complex conjugate pairs and eigenvectors are not simultaneous eigenvectors of $S$. The transition between these two regimes occur via the EP. In the following, we consider the EPs of the transfer matrix of a unit cell $\mathbf{T}_{q}(\varepsilon)$.

{\it Band edges as EPs of transfer matrix of unit cell ---} 
The band edges of the system correspond to $k=0,\pm\pi$. So, from Eq.\eqref{band_condition}, we see that the band edges $\varepsilon= \varepsilon_b$ of the system can be obtained via solution of
\begin{equation}
\label{band_edge_condition}
\left[\frac{{\rm Tr}\left(\mathbf{T}_q(\varepsilon_b)\right)}{2}\right]^2 -1=0.
\end{equation}
Next we note that the determinant of $\mathbf{T}^{(\ell)}(\varepsilon)$, and hence the determinant of $\mathbf{T}_{q}(\varepsilon)$ is $1$. Using this, we can write the eigenvalues of $\mathbf{T}_{q}(\varepsilon)$ as $\lambda_{\pm} = \frac{{\rm Tr}\left(\mathbf{T}_q(\varepsilon)\right)}{2} \pm \sqrt{\left[\frac{{\rm Tr}\left(\mathbf{T}_q(\varepsilon)\right)}{2}\right]^2 -1}.$
From Eq. \eqref{band_edge_condition} we immediately see that at every band edge, there is a single eigenvalue, either both $1$ or both $-1$. Thus, every band edge corresponds to an EP of $\mathbf{T}_q(\varepsilon)$. As we discuss below, this leads to universal anomalous transport behavior at every band edge.

{\it Quantum transport and transfer matrices ---}
We connect the site $1$ and the site $N$ of the lattice chain to fermionic baths, which are modelled by an infinite number of fermionic modes. The associated bath spectral functions being $\mathfrak{J}_1 (\omega)$, $\mathfrak{J}_N (\omega)$. At initial time, the baths are considered to be at their respective thermal states with inverse temperature $\beta \rightarrow \infty$, and chemical potentials $\mu$ and $\mu-\delta \mu$, while the system can reside at some arbitrary initial state (see Fig.~\ref{fig:1}). We are interested in the linear response regime where $\delta \mu$ is small. As long as the bath spectral functions are continuous and the band of the bath encompass all the bands of the system, in the long time limit, the system reaches a unique non-equilibrium steady state (NESS) \cite{Dhar_2006}. 

Using non-equilibrium Green's functions (NEGF) and the nearest neighbour nature of the system, at NESS, the zero temperature conductance can be written as \cite{supp,Dhar_2006,Chaudhuri_2010,Jauho_book},
\begin{equation}
\label{conductance}
\mathcal{G}(\mu) =\frac{\mathfrak{J}_1(\mu)\mathfrak{J}_N(\mu)}{2\pi \left|\Delta_{1,N}(\mu)\right|^2}, 
\end{equation} 
where $\Delta_{1,N}(\mu)$ is the determinant of the inverse of the NEGF. Nearest neighbour hoppings make inverse of the NEGF tridiagonal, as a result, $\Delta_{1,N}(\mu)$ can be obtained from the following relation involving the transfer matrix \cite{supp,Dhar_2006,Chaudhuri_2010}
\begin{equation}
\label{Delta_relation}
\hspace*{-10pt}\left(
\begin{array}{c}
\Delta_{1,N}(\mu) \\
\Delta_{2,N}(\mu)
\end{array}
\right)
\hspace*{-3pt}= \hspace*{-3pt}
\left( \hspace*{-3pt}
\begin{array}{cc}
1 & -\mathbf{\Sigma}_{11}(\mu) \\
0 & 1
\end{array}
\hspace*{-3pt}
\right) \left[\mathbf{T}_{q}(\mu)\right]^{n}
\left(
\hspace*{-3pt}
\begin{array}{c}
1 \\
\mathbf{\Sigma}_{NN}(\mu) 
\end{array}
\hspace*{-3pt}
\right),
\end{equation}
where $n=N/q$ is an integer, $\mathbf{\Sigma}_{\ell \ell}(\omega)= \int \frac{d\omega^\prime}{2\pi}\frac{\mathfrak{J}_\ell(\omega^\prime)}{\omega -\omega^\prime} - i\frac{\mathfrak{J}_\ell(\omega)}{2}$, with $\ell=1, N$, and $\Delta_{2,N}(\mu)$ is the determinant of inverse of the NEGF in absence of the first site. We immediately see from Eqs.(\ref{conductance}) and (\ref{Delta_relation}) that the system-size scaling of conductance is completely independent of the type of bath spectral functions and is entirely governed by the nature of the transfer matrix $\mathbf{T}_{q}(\mu)$.

The system-size scaling of conductance specifies the nature of transport.
In normal conductors, resistance, (i.e, inverse of conductance) is proportional to length, such that resistivity is a well-defined property of the conductor. So, the behaviour is $\mathcal{G}(\mu) \sim N^{-1}$ in normal diffusive transport. Departure from this behavior means resistivity is no longer a well-defined property of the material but depends on the system length. This specifies other types of transport. For ballistic transport, conductance in independent of system length, $\mathcal{G}(\mu) \sim N^{0}$. If  $\mathcal{G}(\mu) \sim N^{-\delta}$, $\delta \neq 0,1$, transport is said to be anomalous. For $0<\delta<1$, transport is called superdiffusive, while for $\delta>1$ transport is called subdiffusive. Apart from these behaviors, conductance can decay exponentially with system length, $\mathcal{G}(\mu) \sim e^{-N/\xi}$, which shows that there is lack of transport beyond a length scale $\xi$.  This behavior is seen in localized systems in presence of disordered or quasiperiodic potentials, with $\xi$ being the localization length. We remark that, for anomalous transport, this classification of transport behavior does not necessarily correspond to classification of transport via spread of density correlations in an isolated system, and may lead to different results \cite{Archak_AAH_2018,Archak_AAH_2019}.

{\it Universal subdiffusive scaling and dissipative quantum phase transition at every band edge---}
The most remarkable result that directly follows from all of the above discussion pertains to the case where $\mu$ is equal to a band edge $\varepsilon_b$ of the system (i.e., $|{\rm Tr}\left(\mathbf{T}_q(\mu)\right)|=2$). As already noted before, the band edges of the system correspond to the EPs of the transfer matrix of a unit cell, both eigenvalues being $1$. Consequently, $\mathbf{T}_q(\mu)$ cannot be diagonalized, but can be taken to the Jordan normal form via a similarity transform, $\mathbf{R}\mathbf{T}_q(\mu)\mathbf{R}^{-1}=\mathbb{I}_2 + (\mathbf{\sigma}_x + i\mathbf{\sigma}_y)/2$. Using properties of Pauli matrices, one then has at $\mu=\varepsilon_b$,
$\left[\mathbf{T}_q(\mu) \right]^n = \mathbf{R}^{-1} \left[\mathbb{I}_2 + n \frac{\mathbf{\sigma}_x + i\mathbf{\sigma}_y}{2} \right]\mathbf{R}$. Note that we do not need the explicit form of $\mathbf{R}$ to obtain this result.
Using this in Eq.(\ref{Delta_relation}), gives $\Delta_{1,N} \sim N$ for $N\gg 1$, and hence, from Eq.(\ref{conductance}), we immediately find $\mathcal{G}(\mu) \sim N^{-2}$. Thus remarkably, because transfer matrix of a unit cell has exceptional points at every band edge, there is a  universal subdiffusive scaling of conductance with system size, with a scaling exponent $2$. 

\begin{figure}
    \centering
    \includegraphics[width=\columnwidth]{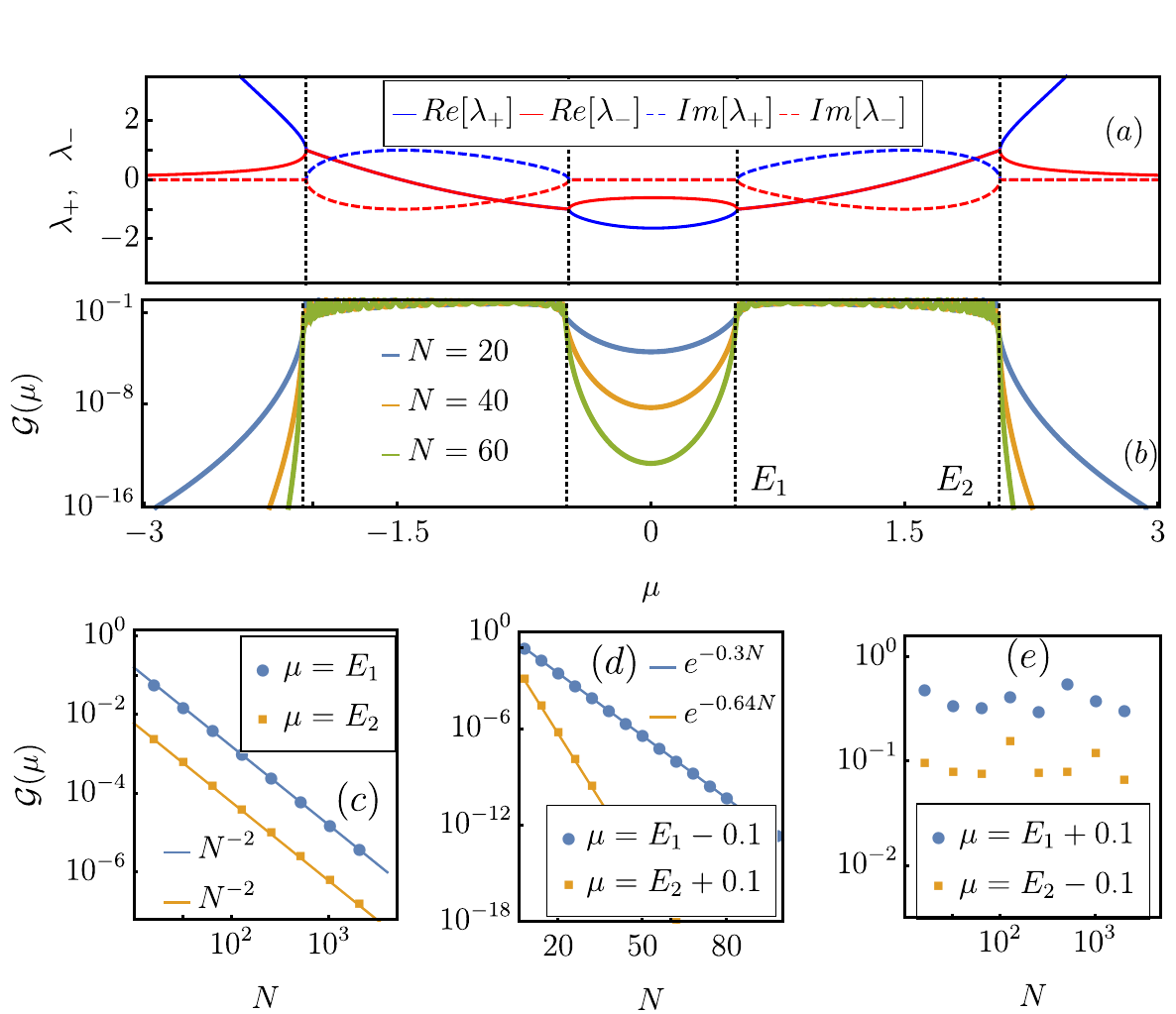}
    \caption{(a) Real and imaginary parts of eigenvalues of $\mathbf{T}_q(\mu)$, given in Eq.(\ref{def_T}), are plotted as a function of $\mu$ for a two-band case, i.e, $q=2$ ($\varepsilon_{l}=\pm 0.5$). The vertical lines correspond to band edges [solution to Eq.~(\ref{band_edge_condition})]. The transition across EP at each band edge is clear. (b) The zero temperature conductance $\mathcal{G}(\mu)$ vs $\mu$ is shown for three different system sizes. The non-analytic change in system-size scaling at every band edge is clear. (c) The universal subdiffusive scaling, $\mathcal{G}(\mu)\sim N^{-2}$ is shown for $\mu$ at two chosen band edges, pointed out in panel (b). (d) The exponential decay of $\mathcal{G}(\mu)$ with $N$ is shown for $\mu$ slightly outside the chosen band. The exponents in the fits are obtained from the formula for localization length. (e) The $\mathcal{G}(\mu) \sim N^0$ behavior is shown for $\mu$ slightly inside the chosen band. All energies are in units of system hopping strength. } 
    \label{fig:2}
\end{figure}

If $\mu$ is within the bands of the chain, then using Eq.(\ref{band_condition}) it can be shown that the eigenvalues of $\mathbf{T}_{q}(\mu)$ are $\lambda_{\pm} = e^{\pm i k}$.  This corresponds to the $S$-symmetry-broken regime of $\mathbf{T}_{q}(\mu)$.  The  $\left[\mathbf{T}_{q}(\mu)\right]^{n}$ 
therefore yields an oscillatory behavior of $\Delta_{1,N}$ with $N$. Thus, within the bands of the chain, $\mathcal{G}(\mu)$ does not show any scaling with $N$, implying ballistic behavior. 

On the other hand, when $\mu$ is outside the band edges of the chain, there is no solution for Eq.(\ref{band_condition}) unless $k$ is purely imaginary ($\kappa \equiv i k$). 
Consequently, the eigenvalues of $\mathbf{T}_q(\mu)$ are real, and therefore corresponds to the $S$-symmetric regime. The eigenvalues can be written as $\lambda_{\pm}=e^{\pm \kappa}$, where ${\rm Tr}\left(\mathbf{T}_q(\mu)\right)=2 \cosh \kappa$. Since one of the eigenvalues of $\mathbf{T}_q(\mu)$ is guaranteed to have magnitude greater than $1$,
$\Delta_{1,N}$ diverges exponentially with system size. Consequently, $\mathcal{G}(\mu) \sim e^{-N/\xi}$, which shows lack of transport beyond a length scale $\xi$. The expression for $\xi$ can be obtained as, 
$\xi^{-1} = \frac{2}{q} \log\left[\left|\frac{{\rm Tr}\left(\mathbf{T}_q(\mu)\right)}{2}\right| + \sqrt{\left[\frac{{\rm Tr}\left(\mathbf{T}_q(\mu)\right)}{2}\right]^2 -1}\right]$,
for  $|{\rm Tr}\left(\mathbf{T}_q(\mu)\right)|>2$.
This behavior is exactly analogous to that observed in a localized disordered or quasiperiodic system, with $\xi$ playing the role of the localization length. In disordered or quasiperiodic systems, the finiteness of the so-called Lyapunov exponent associated with the transfer matrix is taken as the signature of localization \cite{Last_1993,Avila_2015,Avila_2017,Wang_2020,Wang_2021}. For our set-up, this quantity is given by $\ell(\mu)=\lim_{n\rightarrow \infty} \frac{1}{n}  \log\left(\left | \left| \left[\mathbf{T}_q(\mu) \right]^n \right| \right|\right)$,  where $\left| \left| \mathbf{A}\right| \right|$ is the norm of the matrix $\mathbf{A}$. Since one of the eigenvalues of $\left[\mathbf{T}_q(\mu) \right]^n$ diverge exponentially with $n$, the Lyapunov exponent is indeed finite for $\mu$ outside the system bands, and is proportional to $\xi^{-1}$. 

Our main results are given in the table in Fig.~\ref{fig:1}. In the entire discussion above, the nature of the periodic on-site potential $\varepsilon_\ell$,  and its period $q$, which controls the number of bands, is completely arbitrary. This behavior is therefore completely independent of these details. As an example, we demonstrate the two-band case in Fig.~\ref{fig:2} (see \cite{supp} for some other examples).  


The NESS of the chain thus changes non-analytically as a function of $\mu$ at every band edge at zero temperature. This behavior is seen in the large system-size limit, and is completely independent of the nature of bath spectral functions, as well as the strength of system-bath couplings, as long as the steady state is unique. Therefore at every band edge there occurs a dissipative quantum phase transition as a function of $\mu$, which, in our set-up, is not a Hamiltonian parameter but a thermodynamic parameter of the baths. This is unlike most other examples  of dissipative phase transitions (for example, \cite{Rodriguez_2017,Heugel_2019,
Fink_2017,Fitzpatrick_2017,
Jo_2021,Gamayun_2021,Zamora_2020,Jo_2019,Carollo_2019,Minganti_2018,
Marcuzzi_2016, Dagvadorj_2015,Nagy_2015,Bastidas_2012,Prosen_2008}) where a Hamiltonian parameter is changed. Like standard quantum phase transitions, this phase transition occurs strictly at zero temperature, while at finite, but low temperatures, it can be shown to manifest as a finite size crossover \cite{supp}.

Although the transition is independent of the strength of system-bath couplings, the presence of the baths is crucial. This is rooted in the fact that the mechanism for NESS transport relies not only on the chain energy states but also on the energy states available in the baths. The current-current correlations (or the associated density-density correlations) computed in absence of the baths, as often done in the Green-Kubo formalism, will neither show any subdiffusive behavior for $\mu$ at band edges, nor show the existence of a well-defined localization length for $\mu$ outside system bands.  In absence of the baths, in either case, no transport is possible because all bands are either completely full or completely empty. In presence of the baths, even if there is no single particle energy eigenstate for the chain at a chosen value of $\mu$, due to quantum nature of the particles, a small but finite probability exists for few particles to tunnel into and out of the chain, thereby making transport possible. This, in turn, leads to the exotic dissipative phase transition at every band edge. Therefore the non-analytic change in conductance at every band-edge has no obvious analog either in isolated quantum systems or in classical stochastic open systems.


{\it Similarities and differences between band-edges and mobility-edges ---}
The sharp transition as a function of $\mu$ from a regime with a well-defined localization length to a regime of ballistic transport via a `critical point' showing sub-diffusive scaling  is akin to localization-delocalization transitions as a function of energy seen in some quasiperiodic systems (for example, \cite{DasSarma_1988,Luschen_2018,Ganeshan_2015,Li_2017, Wang_2020, Wang_2021,Archak_GAAH_2017}). The energy where this transition happens is called the mobility-edge. In this sense, in a two-terminal set-up, every  band-edge behaves like a mobility-edge. A mobility edge in a two-terminal set-up acts as an energy filter for quantum transport. This property can find potential applications in devising efficient autonomous thermal machines \cite{Chiaracane_2020,Benenti_2017}. Since every band-edge has the same effect, band-edges can also be potentially utilized for the same purpose.  

Despite the analogy between our setup and quasiperiodic systems, there are stark differences. Unlike our setup, for the quasiperiodic systems with mobility edge, the transition in conductance scaling with system size in presence of baths can be linked to a transition in nature of single-particle eigenstates of the system in absence of the baths (see, for example, \cite{DasSarma_1988,Luschen_2018,Ganeshan_2015,Li_2017, Wang_2020, Wang_2021,Archak_GAAH_2017}). This is very different from the transition observed in our set-up which is rooted in transition across EP of the transfer matrix of a unit cell. Note that, while a unit cell is well-defined for a periodic on-site potential, for quasiperiodic on-site potential, a unit cell does not exist.

{\it Conclusions and outlook---}
 
We have shown how non-Hermitian transitions in the transfer matrix of Hermitian Hamiltonians affect the nature of quantum transport (see Fig.~\ref{fig:1}). In doing so, we have united two seemingly disparate concepts: (i) the symmetries and transitions in non-Hermitian matrices studied in non-Hermitian optics, and (ii)  band-structure and quantum transport in fermionic systems studied in condensed matter and statistical physics.  We find that this connection offers an extremely simple way to understand several non-trivial features of mobility edges, localization length and anomalous transport in a two-terminal open system setting, without considering disordered or quasiperiodic potentials. 
We discover a completely different way subdiffusive scaling of conductance, with a universal exponent, can originate: from exceptional points of transfer matrices. We remark that, explaining the origin of sub-diffusive scaling exponents is often a difficult problem \cite{De_Roeck2020,Znidaric_2016,Archak_GAAH_2017}.


Our results pave the way for understanding band-structure and quantum transport in more exotic cases, such as higher dimensional short-ranged systems, in terms of the non-Hermitian properties of their associated transfer matrices \cite{Molinari_1997,Molinari_1998}. But, the transfer matrix picture presented here does not hold in presence of long-range hopping. However, interestingly, the sub-diffusive behavior $\mathcal{G}(\mu) \sim N^{-2}$ at band edges, was also recently numerically seen in presence of long-range, power-law-decaying, hopping \cite{Long_range_subdiffusive}. This points to the super-universality of this behavior at band edges, a deeper understanding of which requires further work.  Another interesting but challenging question is whether the analogy between mobility edges and band edges holds in presence of many-body interactions. Generalization to bosonic transport also remain to be explored.  Investigations in these directions will be carried out in future works.

\begin{acknowledgments}
M. S. acknowledge financial support through National Postdoctoral Fellowship (NPDF), SERB file no.~PDF/2020/000992. B. K. A. acknowledges the MATRICS grant (MTR/2020/000472) from SERB, Government of India and the Shastri Indo-Canadian Institute for providing financial support for this research work in the form of a Shastri Institutional Collaborative Research Grant (SICRG). MK would like to acknowledge support from the project 6004-1 of the Indo-French Centre for the Promotion of Advanced Research (IFCPAR), Ramanujan Fellowship (SB/S2/RJN-114/2016), SERB Early Career Research Award (ECR/2018/002085) and SERB Matrics Grant (MTR/2019/001101) from the Science and Engineering Research Board (SERB), Department of Science and Technology, Government of India. MK acknowledges support of the Department of Atomic Energy, Government of India, under Project No. RTI4001. A.P acknowledges funding from the European Union’s Horizon 2020 research and innovation programme under the Marie Sklodowska-Curie Grant Agreement No. 890884. A.P also acknowledges funding from the Danish National Research Foundation through the Center of Excellence ``CCQ'' (Grant agreement no.: DNRF156).
\end{acknowledgments}

\appendix

\section*{Appendix}
\section*{Zero temperature conductance from NEGF}
\label{NEGF}
As mentioned the main text, we connect the first site of the system and the last site of the system to two fermionic baths, which are modelled by an infinite number of fermionic modes. The full Hamiltonian of the set-up is $\hat{H}= \hat{H}_S + \sum_{\ell=1,N}\hat{H}_{SB_\ell}+\sum_{\ell=1,N}\hat{H}_{B_\ell}$, where 
\begin{align}
   &\hat{H}_{SB_\ell} = \sum_{r=1}^\infty \kappa_{r \ell} \hat{c}_\ell^\dagger \hat{B}_{r \ell} + \kappa_{r \ell}^* \hat{B}_{r \ell}^\dagger \hat{c}_\ell, ~~\hat{H}_{B_\ell} = \sum_{r=1}^\infty \Omega_{r \ell} \hat{B}_{r \ell}^\dagger \hat{B}_{r \ell}. 
\end{align}
 Here, $\hat{B}_{r \ell}$ is the fermionic annihilation operator of the $r$th mode of the bath attached at $\ell$th site of the system, $\Omega_{r \ell}$ is the energy of the same, and $\kappa_{r \ell}$ is the complex coupling between that mode and the system.
The bath spectral functions are defined as 
\begin{align}
    \mathfrak{J}_\ell (\omega) = 2\pi \sum_{r=1}^\infty |\kappa_{r \ell}|^2 \delta(\omega - \Omega_{r \ell}).
\end{align}
As mentioned in the main text, if the bath spectral functions are continuous and the band of the bath encompass the all the bands of the system, in the long time limit, the system reaches a unique NESS.

The NESS can be described in terms of the non-equilibrium Green's function (NEGF). The NEGF of the set-up is the $N \times N$ matrix given by $\mathbf{G}= \mathbf{M}^{-1}(\omega)$, $\mathbf{M}(\omega)=\omega \mathbb{I} - \mathbf{H}-\mathbf{\Sigma}(\omega)$. Here, $\mathbb{I}$ is the $N$ dimensional identity and $\mathbf{H}$ the single-particle Hamiltonian which in our case is a tridiagonal matrix with the diagonal elements being the on-site potential, and the off-diagonal elements being the hopping strength. The $\mathbf{\Sigma}(\omega)$ is  the $N\times N$ diagonal self-energy matrix due to the presence of the baths with non-zero elements given by 
\begin{align}
    \mathbf{\Sigma}_{\ell \ell}(\omega)= \int \frac{d\omega^\prime}{2\pi}\frac{\mathfrak{J}_\ell(\omega^\prime)}{\omega -\omega^\prime} - i\frac{\mathfrak{J}_\ell(\omega)}{2},
\end{align}
 where $\ell$ corresponds to sites where the baths are attached. 

The conductance at NESS can be obtained in terms of the NEGF as
\begin{align}
\label{conductance_finite_temp}
\mathcal{G}(\mu) = \int \frac{d\omega}{2 \pi} \mathcal{T}_{1N}(\omega)\left[-\frac{\partial \mathfrak{n}(\omega)}{\partial \omega}\right],
\end{align}
 with $\mathfrak{n} (\omega)=[e^{\beta(\omega-\mu)}+1]^{-1}$ being the Fermi distribution, and the transmission function between site $1$ and $N$ being $\mathcal{T}_{1N}(\omega)=\mathfrak{J}_1(\omega)\mathfrak{J}_N(\omega)|\mathbf{G}_{1N}(\omega)|^2$, where $\mathbf{G}_{1N}(\omega)$ is the ($1$st,$N$th) element of the NEGF. Since $\mathbf{G}(\omega)$ is the inverse of a tridiagonal matrix with the off-diagonal elements equal to $1$, we have
$
|\mathbf{G}_{1N}(\omega)|^2 = \left|\Delta_{1,N}(\omega)\right|^{-2},~\Delta_{1,N}(\omega) = {\rm det}\left[ \mathbf{M}(\omega) \right]. 
$
We are interested in the zero temperature case, $\beta\rightarrow \infty$. In this case, the conductance is given by
\begin{align}
\label{conductance_supp_mat}
\mathcal{G}(\mu) = \frac{\mathcal{T}_{1N}(\mu)}{2\pi}=\frac{\mathfrak{J}_1(\mu)\mathfrak{J}_N(\mu)}{2\pi \left|\Delta_{1,N}(\mu)\right|^2}. 
\end{align} 
It is important to note that this result crucially depends on the zero temperature nature of Fermi distribution functions, and that the transition at every band-edge happens in $\mathcal{T}_{1N}(\mu)$. Thus, neither for classical systems (for example at high temperature limit), nor for bosonic baths the transition in conductance described in the main text is possible to observe.  However, $\mathcal{T}_{1N}(\omega) \propto |\mathbf{G}_{1N}(\omega)|^2$, and the NEGF is independent of the Fermi distribution of the baths. The NEGF will thus be same for bosonic baths, and also in the high temperature limit. If an observable which is proportional to $|\mathbf{G}_{1N}(\omega)|^2$ can be found under such situations, a similar transition can be seen. But that observable will be different from conductance.

\section*{Transfer matrix from discrete Schr{\"o}dinger equation}
In our context for nearest neighbour hopping model, the transfer matrix connects the amplitude of single particle wave-function between two consecutive sites. The discrete Schr{\"o}dinger equation for nearest neighbour tight-binding model $H \psi= \omega \psi$ reads as, 
\begin{align}
\label{discretescro}
\omega \psi_{\ell}=\varepsilon_{\ell} \psi_{\ell} + \psi_{\ell+1} + \psi_{\ell-1}
\end{align}
Here $\varepsilon_{\ell}$ is the on-site potential energy for $\ell$th site , nearest neighbour hopping strength is 1 and $\psi_{\ell}$ is the amplitude of wave-function at $\ell$th site. We can again rewrite Eq.~\ref{discretescro},
\begin{align}
\label{discretescro2}
\psi_{\ell+1}=(\omega-\varepsilon_{\ell}) \psi_{\ell} -\psi_{\ell-1}    
\end{align}
Now, using Eq.~\ref{discretescro2}, we can write how the amplitude of wave-function at $(\ell+1)$ and $\ell$ th sites are connected with $\ell$ and $(\ell-1)$th site.
\begin{align}
\begin{pmatrix} \psi_{\ell+1} \\ \psi_{\ell}   \end{pmatrix}&= \begin{pmatrix} \omega - \varepsilon_{\ell} & -1 \\
1 & 0\end{pmatrix} \begin{pmatrix} \psi_{\ell} \\ \psi_{\ell-1}   \end{pmatrix} \\ \nonumber
&= \mathbf{T}^{(\ell)}(\omega) \begin{pmatrix} \psi_{\ell} \\ \psi_{\ell-1}   \end{pmatrix}. 
\end{align}
Here $\mathbf{T}^{(\ell)}(\omega)$ is the connecting matrix or the transfer matrix for the $\ell$th site. By knowing this, we could construct the transfer matrix of the unit cell as
\begin{equation}
\mathbf{T}_q(\omega)=\prod_{\ell=1}^q \mathbf{T}^{(\ell)}(\omega),~\mathbf{T}^{(\ell)}(\omega)=\frac{\omega-\varepsilon_{\ell}}{2}\left(\mathbb{I}_2+\sigma_z\right)-i\sigma_y.
\end{equation} 
where $q$ is the periodicity of the lattice.

\section*{Transfer matrix approach to calculate the determinant of tri-diagonal matrices}
In this section, we provide the derivation for Eq.~(6) of the main text. Recall that, the zero temperature conductance in Eq.~(5) of the main text, is related to the determinant of the inverse of NEGF and is given by  
\begin{equation}
\Delta_{1,N}(\omega) = {\rm det}\Big[ \mathbf{M}(\omega) \Big].
\end{equation}
Here $\mathbf{M}(\omega)$ is a tri-diagonal matrix and $\Delta_{i,N}(\omega)$ is the determinant of sub-matrix $\mathbf{M}(\omega)$ starting with the $i$th row and column
and ending with $N$th row and column. For a tight-binding model with  on-site potential energy $\varepsilon_i$ for $i$th site, $\mathbf{M}(\omega)$ has the form,
\begin{widetext}
\begin{align}
\mathbf{M}(\omega)=\begin{pmatrix} \omega-\mathbf{\Sigma_{11}} -\varepsilon_1 & -1 & 0 & 0 & 0 & \ldots\\
-1 & \omega - \varepsilon_2 & -1 & 0 & 0 & \ldots \\
0 & -1 & \omega & -1 & 0 &\ldots \\
\vdots & \vdots & \vdots & \vdots & \vdots & \vdots \\
0 & 0 & 0 & -1 & \omega & -1 \\
0 & 0 & 0 & 0 & -1 & \omega - \mathbf{\Sigma_{NN}}-\varepsilon_N
\end{pmatrix},
\end{align}
\end{widetext}
where for brevity, we have suppresed the frequency dependence of the self-energies.
To calculate the determinant of $\mathbf{M}(\omega)$, the iterative equations are,
\begin{widetext}
\begin{align}
\label{iteration}
\Delta_{1,N}(\omega)&=(\omega - \mathbf{\Sigma_{11}}-\varepsilon_1) \Delta_{2,N}(\omega) -\Delta_{3,N}(\omega) \\ \nonumber
\Delta_{i,N}(\omega)&=(\omega - \varepsilon_i) \Delta_{i+1,N}(\omega)-\Delta_{i+2,N}(\omega),~~2\leq i \leq N-2 \\ \nonumber
\Delta_{N-1,N}(\omega)&=(\omega - \varepsilon_{N-1}) \Delta_{N,N}(\omega)-1 \\ \nonumber
\Delta_{N,N}&=\omega-\mathbf{\Sigma_{NN}}-\varepsilon_N
\end{align}
Using these Eq.~\ref{iteration}, we can write,
\begin{align}
\label{equation6}
\begin{pmatrix}
\Delta_{1,N} \\
\Delta_{2,N}
\end{pmatrix}&=\begin{pmatrix}
\omega - \mathbf{\Sigma_{11}} -\varepsilon_1 & -1 \\
1 & 0
\end{pmatrix} \begin{pmatrix}
\Delta_{2,N} \\
\Delta_{3,N}
\end{pmatrix}    \\ \nonumber
&=\begin{pmatrix}
\omega - \mathbf{\Sigma_{11}} - \varepsilon_1 & -1 \\
1 & 0
\end{pmatrix}  \begin{pmatrix}
\omega -\varepsilon_2  & -1 \\
1 & 0
\end{pmatrix} \begin{pmatrix}
\omega -\varepsilon_3  & -1 \\
1 & 0
\end{pmatrix} \ldots \begin{pmatrix}
\omega -\varepsilon_{N-1}  & -1 \\
1 & 0
\end{pmatrix} \begin{pmatrix}
\Delta_{N,N} \\
1
\end{pmatrix} \\ \nonumber
&=\begin{pmatrix}
\omega - \mathbf{\Sigma_{11}} -\varepsilon_1 & -1 \\
1 & 0
\end{pmatrix} \begin{pmatrix}
\omega -\varepsilon_2  & -1 \\
1 & 0
\end{pmatrix} \begin{pmatrix}
\omega -\varepsilon_3  & -1 \\
1 & 0
\end{pmatrix} \ldots \begin{pmatrix}
\omega -\varepsilon_{N-1}  & -1 \\
1 & 0
\end{pmatrix}  \begin{pmatrix}
\omega - \mathbf{\Sigma_{NN}} - \varepsilon_N & -1 \\
1 & 0
\end{pmatrix}  \begin{pmatrix}
1 \\
0
\end{pmatrix}  \\ \nonumber
&=\begin{pmatrix}
\omega - \mathbf{\Sigma_{11}} -\varepsilon_1 & -1 \\
1 & 0
\end{pmatrix} \begin{pmatrix}
\omega-\varepsilon_1  & -1 \\
1 & 0
\end{pmatrix}^{-1} \begin{pmatrix}
\omega -\varepsilon_1  & -1 \\
1 & 0
\end{pmatrix}  \ldots \begin{pmatrix}
\omega -\varepsilon_{N}  & -1 \\
1 & 0
\end{pmatrix} \begin{pmatrix}
\omega-\varepsilon_N  & -1 \\
1 & 0
\end{pmatrix}^{-1} 
\begin{pmatrix}
\omega - \mathbf{\Sigma_{NN}} -\varepsilon_N \\
1 
\end{pmatrix}   \\ \nonumber
&=\begin{pmatrix}
1 & -\mathbf{\Sigma_{11}} \\
0 & 1
\end{pmatrix} [\mathbf{T}_q(\omega)]^n \begin{pmatrix}
1 \\
\mathbf{\Sigma_{NN}}
\end{pmatrix}.
\end{align} 
\end{widetext}
With periodic on-site potential $\varepsilon_{\ell}=\varepsilon_{\ell +q}$, $\mathbf{T}_q(\omega)$ is the transfer matrix for the unit cell and $n=N/q$ is the number of unit cells. 


\begin{figure}
    \centering
    \includegraphics[width=\columnwidth]{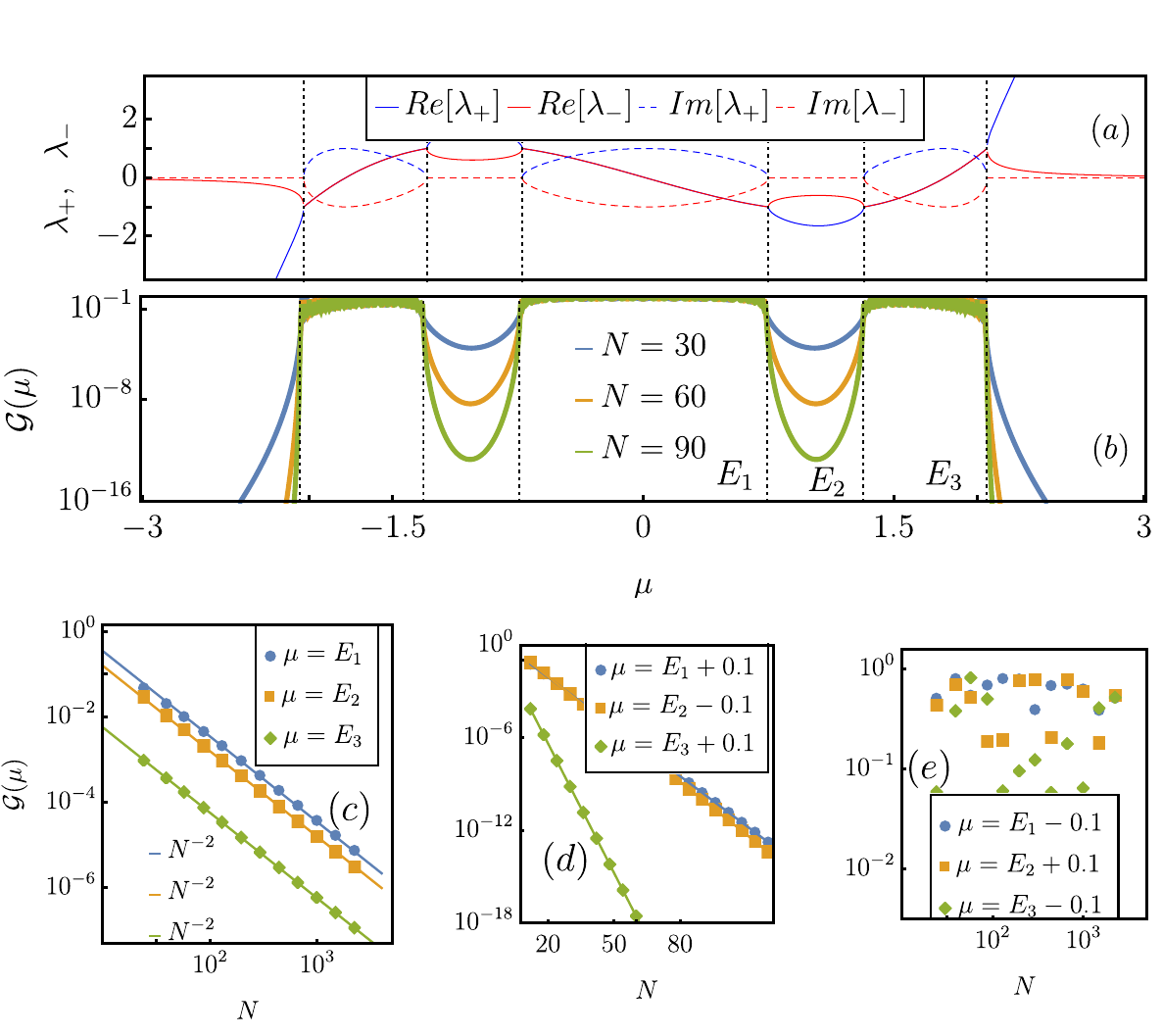}
    \caption{(a) Real and imaginary parts of eigenvalues of $\mathbf{T}_{q}(\mu)$,are plotted as a function of $\mu$ for a three-band case, i.e, $q=3$ ($\varepsilon_{l}=\pm 0.5,~0$). The vertical lines correspond to band edges. The transition across EP at each band edge is clear. (b) The zero temperature conductance $\mathcal{G}(\mu)$ vs $\mu$ is shown for three different system sizes. The non-analytic change in system-size scaling at every band edge is clear. (c) The universal subdiffusive scaling, $\mathcal{G}(\mu)\sim N^{-2}$ is shown for $\mu$ at two chosen band edges, pointed out in panel (b). (d) The exponential decay of $\mathcal{G}(\mu)$ with $N$ is shown for $\mu$ slightly outside the chosen band.  (e) The $\mathcal{G}(\mu) \sim N^0$ behavior is shown for $\mu$ slightly inside the chosen band. All energies are in units of system hopping strength. } 
    \label{fig:3}
\end{figure}

\section*{Transformation to Jordan normal form at every band-edge}
Even if a matrix is not diagonalizable, it can always be taken to the Jordan normal form via a similarity transform. Any $2\times 2$ matrix which is not proportional to identity becomes non-diagonalizable if its characteristic polynomial has a single solution, so that there is only one eigenvalue, say $\lambda$. In such cases, the Jordan normal form is
\begin{align}
\left(
\begin{array}{cc}
\lambda & 1 \\
0 & \lambda
\end{array}
\right).
\end{align}
Every band-edge $\mu=\varepsilon_b$, corresponds to an exceptional point of the transfer matrix of a unit cell, with only one eigenvalue $\lambda=1$. So, $\mathbf{T}_q(\mu=\varepsilon_b)$ is not diagonalizable. Correspondingly, we can transform it into the Jordan normal form and write,
\begin{align}
\mathbf{R}(\mu)\mathbf{T}_q(\mu)\mathbf{R}^{-1}(\mu)&= \left(
\begin{array}{cc}
1 & 1 \\
0 & 1
\end{array}
\right)=\mathbb{I}_2 + \frac{\mathbf{\sigma}_x + i\mathbf{\sigma}_y}{2}\\ \nonumber 
~\Rightarrow \mathbf{T}_q(\mu)& =\mathbf{R}^{-1}(\mu) \left[ \mathbb{I}_2 + \frac{\mathbf{\sigma}_x + i\mathbf{\sigma}_y}{2} \right] \mathbf{R}(\mu),
\end{align}
where $\mathbf{\sigma}_{x,y}$ are the corresponding Pauli matrices.
Conductance is governed by the $n$th power of $\mathbf{T}_q(\mu)$,
\begin{align}
\left[\mathbf{T}_q(\mu)\right]^n = \mathbf{R}^{-1}(\mu) \left[ \mathbb{I}_2 + \frac{\mathbf{\sigma}_x + i\mathbf{\sigma}_y}{2} \right]^n \mathbf{R}(\mu).
\end{align}
Noting that 
\begin{align}
\left[\frac{\mathbf{\sigma}_x + i\mathbf{\sigma}_y}{2} \right]^n = 0,~~\forall~~n>1,
\end{align}
and performing a binomial expansion, we see that
\begin{align}
\left[\mathbf{T}_q(\mu)\right]^n = \mathbf{R}^{-1}(\mu) \left[ \mathbb{I}_2 + n\frac{\mathbf{\sigma}_x + i\mathbf{\sigma}_y}{2} \right] \mathbf{R}(\mu).
\end{align}
This shows that at every band edge, $\left[\mathbf{T}_q(\mu)\right]^n \sim n$.
We do not need the explicit form of $\mathbf{R}(\mu)$ for this result. In fact, writing down the equations for determining the elements of $\mathbf{R}(\mu)$, one finds that the elements are not uniquely determined. We show this next explicitly for the one-band case, where the on-site potential $\varepsilon_\ell=0$.

At the band-edge of the one-band model ($\mu=\pm 2$), the transfer matrix has the form $ \mathbf{T}_q (\mu=2)=\begin{pmatrix}
2 & -1 \\
1 & 0
\end{pmatrix}$ or $\mathbf{T}_{q}(\mu\!=\!-2)=\begin{pmatrix}
-2 & -1 \\
1 & 0
\end{pmatrix}$. As these matrices are not diagonalizable with a transformation, we can take it to the Jordan-normal form. Thus, $\mathbf{R}^{-1}(2) \begin{pmatrix}
1 & 1 \\
0 & 1
\end{pmatrix} \mathbf{R}(2) = \begin{pmatrix}
2 & -1 \\
1 & 0
\end{pmatrix}$ and $\mathbf{R}^{-1}(-2) \begin{pmatrix}
1 & 1 \\
0 & 1
\end{pmatrix} \mathbf{R}(-2) = \begin{pmatrix}
-2 & -1 \\
1 & 0
\end{pmatrix}$. 

Let us now calculate $\mathbf{R}(2)$. Let us consider, $\mathbf{R}(2)=\begin{pmatrix} a & b \\
c & d
\end{pmatrix}$. Now, to know the values of $a$, $b$, $c$ and $d$, we have $4$ equations,
\begin{align}
  \frac{1}{ad-bc} \begin{pmatrix}
  d & -b \\
  -c & a
  \end{pmatrix} \begin{pmatrix}
  1 & 1 \\
  0 & 1
  \end{pmatrix}  \begin{pmatrix}
  a & b \\
  c & d
  \end{pmatrix}=  \begin{pmatrix}
2 & -1 \\
1 & 0
\end{pmatrix} \\ \nonumber
\implies \frac{1}{ad-bc} \begin{pmatrix} ad+ cd -bc & d^2 \\
-c^2 & -bc-cd+ad
\end{pmatrix}= \begin{pmatrix}
2 & -1 \\
1 & 0
\end{pmatrix}.
\end{align}
Fixing $ad-bc=1$, $c=\pm i$ and $d=\pm i$. If we choose, $d=-i$ and $c=i$, we will get a relation between $a$ and $b$ and the relation is $a+b=i$. We can consider any choice of $a$ and $b$ which satisfies this equation. Lets say, we consider $a=i$ and $b=0$, then $\mathbf{R}(2)=\begin{pmatrix}
i & 0 \\
i & -i
\end{pmatrix}$. Thus, $\mathbf{R}(2)$ is not unique. Regardless, the fact that it is possible to find $\mathbf{R}(2)$ is enough to show $\left[\mathbf{T}_q(2)\right]^n \sim n$, which in turn guarantees a sub-diffusive scaling of conductance.

\section*{Subdiffusive behaviour at band edges for three band model}
The non-analytic change in conductance and the sub-diffusive scaling of conductance occurs at every band-edge, irrespective of the nature and period of the on-site potential. Here we explicitly show the same for a three-band model, i.e, where the period of the potential is three sites.

Let's say, $\varepsilon_a$, $\varepsilon_b$ and $\varepsilon_c$ are the on-site potential which is repeating. Thus, the transfer matrix for the unit cell is,
\begin{equation}
\mathbf{T}_{q}(\omega)=\begin{pmatrix}
\omega -\varepsilon_a & -1 \\
1 & 0
\end{pmatrix} \begin{pmatrix}
\omega -\varepsilon_b & -1 \\
1 & 0
\end{pmatrix} \begin{pmatrix}
\omega -\varepsilon_c & -1 \\
1 & 0
\end{pmatrix}. 
\end{equation}
Now, $\mathrm{Tr}[\mathbf{T}_{q}(\omega)]=\pm 2$, will give the $6$ band edges of the system. Now, if we choose $\varepsilon_a=-0.5$, $\varepsilon_b=0.0$ and $\varepsilon_c=0.5$, then the band edges occur at $\pm 2.05505$, $\pm 1.31491$ and $\pm 0.740139$. At this band-edges, $\mathbf{T}_{q}(\omega)$ also has exceptional point. In Fig.~\ref{fig:3}, we have shown the subdiffusive behaviour due to exceptional points when the chemical potential of bath $\mu$ exactly at the band edges.
\section*{Conductance at finite temperature}
\begin{figure}
    \centering
    \includegraphics[width=\columnwidth]{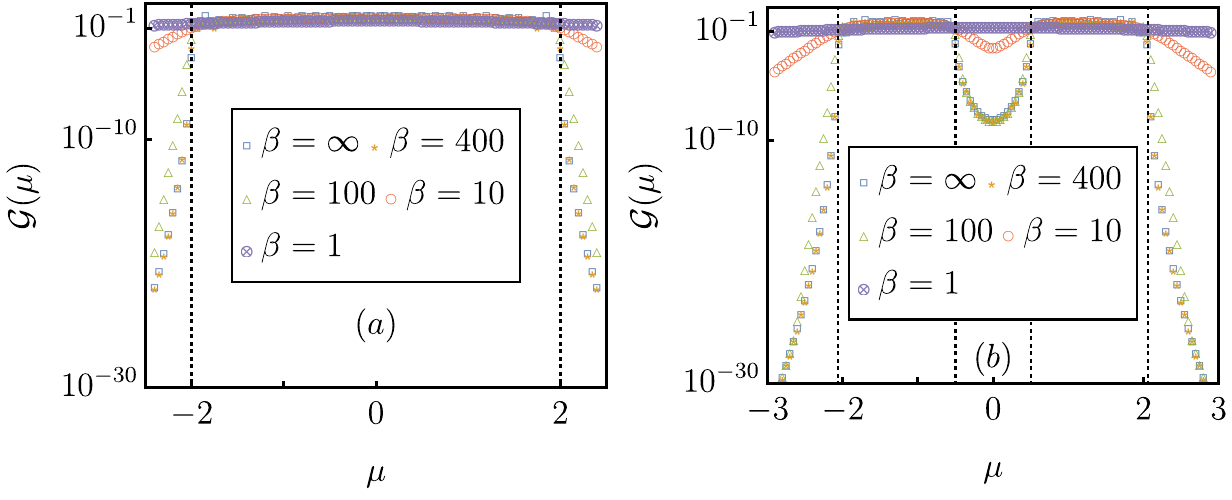}
    \caption{Here, we have plotted the finite temperature conductance for (a) one-band model and (b) two-band model lattice system respectively for system size $N=40$. We can clearly see that the non-analytic changes at the band-edges of the conductance get smoothed at finite temperature. } 
    \label{fig:6}
\end{figure}

The non-analytic change in conductance at every band-edge corresponds to a dissipative quantum phase transition. Like standard quantum phase transitions, this is a strictly zero temperature phenomenon. From Eq.\eqref{conductance_finite_temp}, we see that at finite temperature there will be contributions over the entire range of frequencies. Since for $\omega$ within the system bands $\mathcal{T}_{1N}(\omega)\sim N^0$, while outside system bands and at band-edges it decays with $N$, for large enough $N$, eventually conductance will show ballistic scaling $\mathcal{G}(\mu) \sim N^0$. However, at finite but low temperatures, $-\frac{\partial \mathfrak{n}(\omega)}{\partial \omega}$ is highly peaked at $\mu$. So, at finite but low temperatures, we expect the transition to survive up to a finite size. Thus, like standard quantum phase transitions, at finite but low temperatures, there will be a finite-size crossover, which will be lost on further increasing temperature. In Fig.~\ref{fig:6} we show conductance as a function of $\mu$ at various temperatures for one-band and two-band models, for a system of $N=40$. We clearly see the expected behavior.

\section*{Beyond linear response results}
\begin{figure}
    \centering
    \includegraphics[width=\columnwidth]{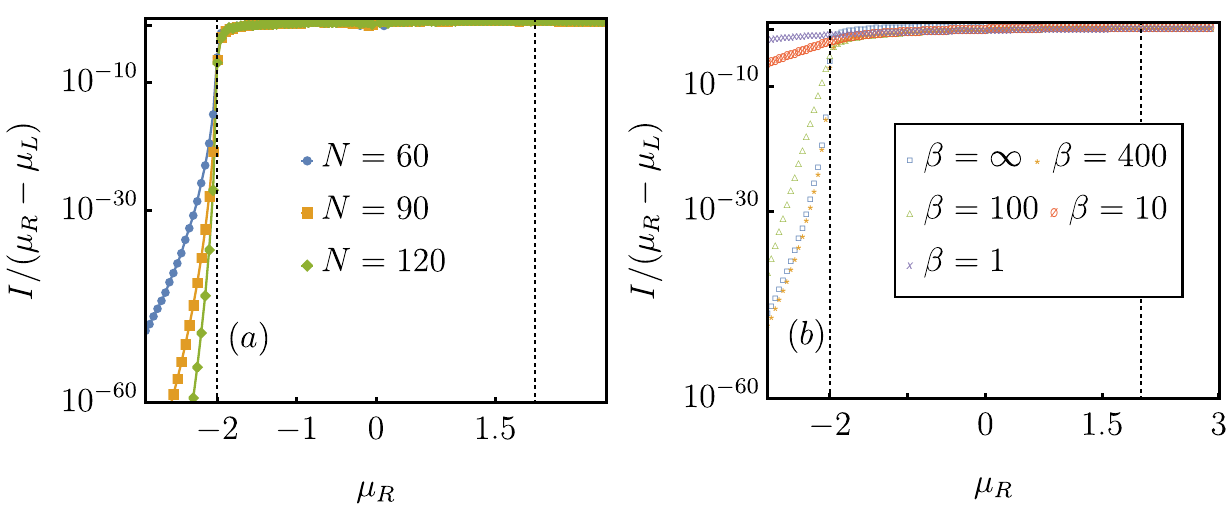}
    \caption{The plot shows how the beyond linear response current $I$ shows a transition when the chemical potential of right bath $\mu_R$ crosses the lower band edge, with $\mu_L$ being kept below the lower band-edge. (a) In the first plot, both the baths are kept at zero temperature and we have shown the crossing for different system sizes. (b) In the second plot, for a fixed system size $N=60$, we have shown the crossing for different temperatures. Both of these plots are for single-band model.} 
    \label{fig:4}
\end{figure}

In this section, we show the effect of exceptional points of transfer matrix beyond linear response. The average charge current in the steady state flowing out of the $N$ th site from $1$ st site is given by the Landauer-B{\"u}ttiker formula,
\begin{align}
\label{current}
I=\int \frac{d\omega}{2\pi}\mathcal{T}_{1N}(\mathfrak{n}_1(\omega)-\mathbf{n}_2(\omega))    
\end{align}
Here, $\mathfrak{n}_1 (\omega)=[e^{\beta(\omega-\mu_L)}+1]^{-1}$ and $\mathfrak{n}_2 (\omega)=[e^{\beta(\omega-\mu_R)}+1]^{-1}$.  In the zero temperature limit, Eq.~\ref{current}, reads as,
\begin{align}
\label{current1}
I=\int\limits_{\mu_L}^{\mu_R} \frac{d\omega}{2\pi}\mathcal{T}_{1N}(\mathfrak{n}_1(\omega)-\mathbf{n}_2(\omega)).    
\end{align}
In Fig.~\ref{fig:4}(a), keeping the temperatures of both the baths zero, and fixing the chemical potential of left bath at $\mu_L=-3.0$, we are changing the chemical potential of the right bath $\mu_R$ for the single band model. With increasing $\mu_R$, we can see a clear transition from exponentially decaying regime to ballistic regime. The transition occurs when the chemical potential of right bath is exactly at the lower band edge $\mu_R=-2.0$. As we have kept chemical potential of one of the baths outside the lower band-edge, we only see the transition around the lower band edge. With finite identical temperature of baths, the sharp transition at lower band-edge gets modified and transition point around the lower band-edge gets smoothed. We have shown this in Fig.~\ref{fig:4}(b).

\bibliography{ref_transfer_matrix_EP}
\end{document}